\documentstyle[prl,aps,multicol,epsf]{revtex}

\newcommand{\Sp}{\mbox{Sp}}
\newcommand{\N}{{\cal N}}
\newcommand{\Ham}{{\cal H}}
\newcommand{\kBT}{k_{\rm B}T}
\newcommand{\av}[2]{\left\langle #2 \right\rangle_{#1}}

\begin{document}

\title{The effect of an external magnetic field on the gas-liquid
transition in the Ising spin fluid}

\author{R.O.Sokolovskii\cite{email}}
\address{Institute for Condensed Matter Physics,
Svientsitskii 1, Lviv 290011, Ukraine. }

\maketitle

\begin{abstract}
The theoretical phase diagrams of the magnetic (Ising) lattice
fluid in an external magnetic field is presented. It is
shown that, depending on the strength of the nonmagnetic interaction
between particles, various effects of external field on the Ising fluid
take place.  In particular, at moderate values of the nonmagnetic
attraction the field effect on the gas-liquid critical temperature is
nonmonotoneous.  A justification of such behavior is given. If
short-range correlations are taken into account (within a cluster
approach), the Curie temperature also depend on the nonmagnetic
interaction.  \end{abstract}

\begin{center}
PACS numbers
		64.70.Fx, 
		77.80.Bh, 
		75.50.Mm, 
	 	64.60.Kw. 
\end{center}

\begin{multicols}{2}

Anisotropic liquids are very sensitive matters. Such are nematic liquid
crystals and ferrofluids. Many efforts have been made in order to
investigate effects of shape and flexibility of the molecules, of long
and short-range interactions on the properties of anisotropic liquids.
External field effects are still worthier of attention, because the
application of an external field allows to change properties of
anisotropic liquids dynamically (in contrast to the effects of molecules'
shape, etc., which are static).

In ferrofluids an
external magnetic field removes the magnetic order-disorder transition,
nevertheless the first order transitions between ferromagnetic phases of
different densities remain.  The external field deforms the phase diagram
of a magnetic fluid shifting coexistence lines between these phases.
Kawasaki studied a magnetic lattice gas \cite{Kawasaki}, which implements
one of the ways to model properties of magnetic fluids. On
temperature-density phase diagrams in Ref.~\cite{Kawasaki} one can see
that the gas-liquid binodal of the magnetic fluid significantly lowers
after the application of an external magnetic field. Vakarchuk with
coworkers \cite{VRP} and Lado et al.  \cite{Lado,LadoPRL} studied
another model of magnetic fluids --- the fluid of hard spheres with
embedded Heisenberg spins. They concluded that in such a system the
temperature of the gas-liquid critical point (the top of binodal)
increases after the application of an external field. The nature of such a
discrepancy can be various, because the types of models (continuum fluids
\cite{VRP,Lado,LadoPRL} and the lattice gas \cite{Kawasaki}) as well as
approximations used differ.

In this letter results of a more detailed investigation of
the magnetic (Ising) lattice gas are reported. The first new point is the
inclusion of the nonmagnetic interaction between particles. Another one
consists in overcoming of limitations applied by the mean field
approximation (MFA).  It is known that this approximation is good for very
long-range potentials only (and becomes accurate for a family of the
infinite-ranged ones, the so-called Kac potentials
\cite{Lebovitz,Balescu}). The MFA can not reproduce some essential
features of the systems with the nearest-neighbor interaction (such as the
percolation phenomena in the quenched diluted Ising model, differences in
magnetic properties of the quenched site-disordered Ising model and the
annealed one \cite{Vaks}).  Such drawbacks can be overcome with the
two-site cluster approximation (TCA) \cite{APP}.

We shall show within both the MFA and the TCA that at different values of
the nonmagnetic interaction between particles the Ising fluid demonstrates
various effects of external fields.  In particular, at moderate values of
the nonmagnetic attraction the field effect on the critical temperature is
nonmonotoneous.

In lattice models of a fluid its
particles are allowed to occupy only those spatial positions which belong
to sites of a chosen lattice. The configurational integral of a simple
fluid is in such a way substituted by the partition function

\begin{eqnarray}
&&Z= \Sp \exp \left( -\beta\Ham  \right),~\beta=1/(\kBT)
,\nonumber\\
&&\Ham=H-\mu N= -\frac12\sum_{ij}I_{ij}n_i n_j-\mu\sum_i n_i
,\nonumber
\end{eqnarray}
where $n_i$, which equals 0 or 1, is a number of particles at site $i$.
$\Sp$ means a summation over all occupation patterns. The total
number $N$ of particles is allowed to fluctuate, $\mu$ is a chemical
potential, which should be determined from the relation

\begin{equation}
N=\av{\Ham}{\sum_i n_i};~
\av{\Ham}{\cdots}=Z^{-1}\Sp(\cdots)\exp(-\beta\Ham)
.
\end{equation}
Lattice models, due to particles can not approach closer than the lattice
spacing allows, automatically preserve the essential feature of molecular
interaction:  nonoverlapping of particles. The lattice fluid with nearest
neighbor interaction is known to demonstrate the gas-liquid transition
only.  Nevertheless, the lattice gas with interacting further neighbors
possesses a realistic (that means, argon-like) phase diagram with all
transitions between the gaseous, liquid and solid phases being present
\cite{HallStell}.

We shall consider a {\em magnetic} fluid in which the particles carry Ising
spins $S_i=\pm1$ and there is also an additional exchange interaction
between the particles

\begin{eqnarray}
H-\mu N &=& -\frac12\sum_{ij}J_{ij}S_in_i S_jn_j - h\sum_i S_in_i
\nonumber\\
        & & -\frac12\sum_{ij}I_{ij}n_i n_j - \mu\sum_i n_i
\label{Ham}
\end{eqnarray}
Each site can be in three states: empty (i), occupied by
the particle with spin up (ii) or down (iii). The trace in the partition
function implies a summation over all $3^\N$ states, where $\N$ is a
number of sites.

An interaction of fluctuations are totally neglected in the mean field
approximation used in the previous studies of the model
\cite{Kawasaki}. This can partially be recovered using the
idea of ``clusters". The partition function of a finite group of particles
in an external field can be evaluated explicitly. A contribution of the
other particles may be expressed in terms of the effective field, and this
field has to be evaluated selfconsistently.  From such a point of view the
MFA is a one-site cluster approximation, in which each cluster comprises
one site. Increasing the size of clusters one may expect to obtain more
accurate results.  Indeed, the results of the two-site cluster
approximation turns to be accurate for the one-dimensional systems
\cite{Balagurov} and on the Cayley tree \cite{Vaks}. Here we shall
formulate such an approximation for the Ising lattice gas with the nearest
neighbor interactions.  For the sake of brevity we shall not use the
cluster expansion formulation which has some advantages, such as the
possibility to calculate corrections of a higher order and correlation
functions of the model \cite{APP}.  Instead we shall rely on the first
order approximation and closely follow the derivation by Vaks and Zein
\cite{Vaks}. Let us introduce the effective-field Hamiltonian of a single
site

\begin{equation}
H_i-\mu n_i\equiv\Ham_i=-\tilde h S_in_i-\tilde \mu n_i
,
\end{equation}
where $\tilde h=h+z\varphi$, $\tilde \mu=\mu+z\psi$, $\varphi$ and $\psi$
are effective fields substituting for interactions with nearest neighbor
sites, $z$ is a first coordination number of the lattice. In the
two-site Hamiltonian the interaction between a pair of the nearest
neighbor sites is taken into account explicitly

\begin{eqnarray}
H_{ij}-\mu n_i-\mu n_j\equiv\Ham_{ij}&=&
        -J_{ij}S_in_i S_jn_j-\tilde h'S_in_i-\tilde h'S_jn_j
\nonumber\\
&& -I_{ij}n_i n_j-\tilde \mu'n_i-\tilde \mu'n_j
,
\end{eqnarray}
where $\tilde h'=h+z'\varphi$, $\tilde \mu'=\mu+z'\psi$, and $z'=z-1$ due
to one of the neighbors is already taken into account.  The fields have
to be found from the selfconsistency conditions that require an equality
of average values calculated with the one-site and two-site Hamiltonians.
To determine $\varphi$ and $\psi$ it is sufficient to impose these
conditions on the average values of spin $m=\av{\Ham}{S_in_i}$ and of the
occupation number $n=\av{\Ham}{n_i}$,

\begin{equation}
\av{\Ham_i}{n_i}=\av{\Ham_{ij}}{n_i};~
\av{\Ham_i}{S_in_i}=\av{\Ham_{ij}}{S_in_i}
\label{selfcon}
.
\end{equation}
This approximation leads to the following expression for
the internal energy

\begin{eqnarray}
U/\N&=&-\frac12 J_0\av{\Ham_{ij}}{S_in_i S_jn_j}
        -h\av{\Ham_{ij}}{S_in_i}
\nonumber\\
&&-\frac12 I_0\av{\Ham_{ij}}{n_i n_j}
\label{U}
,
\end{eqnarray}
where $J_0=\sum_j J_{ij}=zK$ and $I_0=\sum_j I_{ij}=zV$ are integral
interaction strengths, $K$ and $V$ denote, respectively, the magnetic
coupling and the nonmagnetic attraction between nearest neighbor sites.
The expression (\ref{U}) can be computed explicitly in terms of the fields
$\varphi$ and $\psi$ and model parameters.  The other thermodynamic
potentials can be found in a straightforward way.  For example, the grand
thermodynamic potential $\Omega$ of the model satisfies the following
Gibbs-Helmholtz equation

\begin{equation}
\frac{\partial \beta\Omega}{\partial \beta}=U-\mu N
.
\end{equation}
The solution of this differential equation, taking into account relations
(\ref{selfcon}), reads

\begin{equation}
\beta\Omega/\N=z'\ln\Sp\exp(-\beta\Ham_i)
	-\frac z2 \ln\Sp\exp(-\beta\Ham_{ij})
\label{omega}
.
\end{equation}
It is possible to build isotherms of the fluid using the thermodynamic
relation $\Omega=-PV$ and expression (\ref{omega}) and solving the system
of nonlinear
selfconsistency equations (\ref{selfcon}). It is possible and convenient to
exclude the
chemical potential $\mu$ and the field $\psi$ from final equations of
state:

\begin{eqnarray}
\beta PV/\N&=&-\ln(1-n)+\frac z2\ln [(1-n)(1-x)+xr]
\label{PV}
,
\\
\tanh\beta\tilde h
  &=&p    \frac{\sinh2\beta\tilde h'}{\cosh2\beta\tilde h'+\exp(-2\beta K)}
\nonumber\\
  &&+(1-p)\tanh\beta\tilde h'
\label{sc2}
,
\end{eqnarray}
where
\begin{eqnarray}
p&=&1-(1-n)/r,~r=0.5+\sqrt{(n-0.5)^2+n(1-n)/x}
\nonumber,\\
x&=&\frac{2\exp(-\beta V-\beta K)\cosh^2\beta\tilde h'}
        {\cosh2\beta\tilde h'+\exp(-2\beta K)}
\label{diff}
,
\end{eqnarray}
$p$ is a probability that a randomly chosen nearest neighbor site to a
given particle is occupied. In the limit $z\to\infty$ and $K\to0$, $V\to0$
(keeping $J_0$ and $V_0$ constant) TCA formulae
(\ref{PV}--\ref{diff}) turn into the results of the MFA.

At low temperatures the isotherms of the fluid contain the ``liquid" and
``gaseous" parts separated by a region of the negative compressibility.
The thermodynamic states, in which the compressibility of the uniform
fluid is negative ($\frac{\partial P}{\partial n}<0$), constitute the
spinodal region on the temperature-density phase diagram. In this region
the fluid is thermodynamically unstable and must separate on phases of
different densities. The densities of coexisting phases can be determined
with the Maxwell rule of areas applied to the nonmonotoneous sections of
the isotherms.  Also at sufficiently low temperatures and $h=0$ an isotherm
of the model has a break at the density in which nonzero solution of
selfconsistency equation (\ref{sc2}) for the field $\varphi$ appears, and
the second order phase transition to the ferromagnetic phase occurs.

\end{multicols}

\begin{figure*}
\centerline{
 \begin{tabular}{cc}
 \epsfxsize=230pt\epsffile{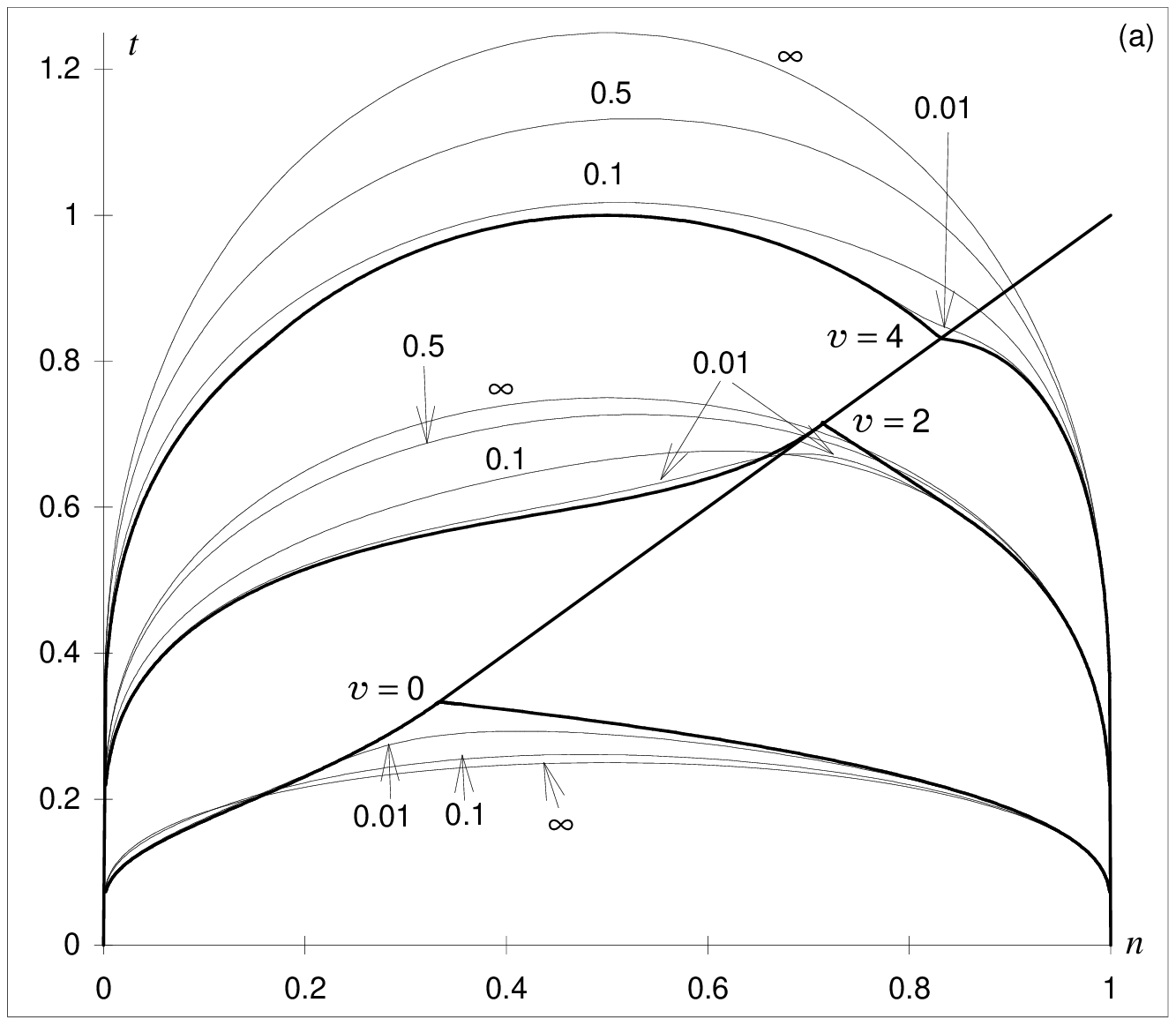}  &  \epsfxsize=230pt\epsffile{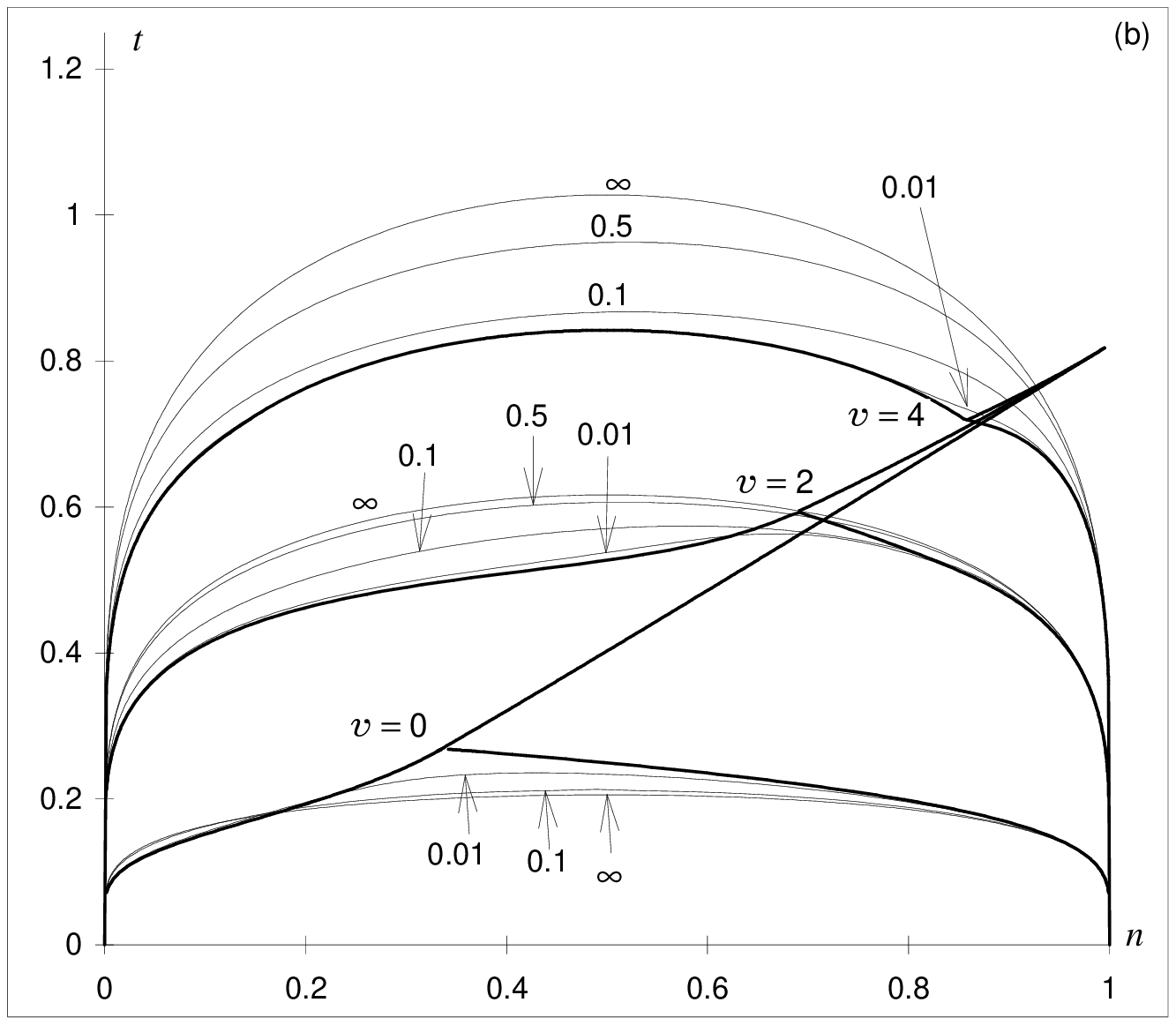}
 \end{tabular}
}
\caption{The phase diagram of the magnetic (Ising) lattice gas
within the mean field (a) and two-site cluster (b) approximations.
$n=N/\N$ and $t=\kBT/J_0$ are reduced density and temperature,
respectively.  Picture b represents TCA results for the model with the
nearest neighbor interactions on the simple cubic lattice ($z=6$). The
results of MFA (picture a) are independent of interaction range and
lattice structure.  Three families of lines are built for three values of
nonmagnetic interaction $v=I_0/J_0$. The thick lines are binodals and
Curie lines at zero external field.  The thin lines with attached numbers
represent the binodals at nonzero values of the external field $h/J_0$.  }
\label{mf&tca} \end{figure*}

\begin{multicols}{2}

Figure \ref{mf&tca}a shows the temperature-density phase diagram of the
model within the mean field approximation at different model parameters.
There are three families of lines for three values of the nonmagnetic
interaction strength $v=I_0/J_0$. The bold lines correspond to zero field
case ($h=0$).  The straight line is a line of Curie points that separates
paramagnetic and ferromagnetic regions.  Within the mean field
approximation the slope of the Curie line is independent of $v$.  Under
binodals (convex lines resting on the points (0,0) and (1,0) on the phase
diagram) a phase separation takes place:  one of the phases (vapor) is
rarefied, the other (liquid) is denser. The nonmagnetic attraction between
particles, of course, favors the phase separation --- the binodal moves
upward with increasing $v$.  At small $v$ the phase diagram possesses of
the tricritical point: the top of binodal lays on the Curie line, the
liquid is ferromagnetic and the gas is paramagnetic. The tricritical point
disappears at large $v$, in this case the top of binodal deviates from the
Curie line in the paramagnetic region.

The external field dissolves the ferromagnetic transition and
therefore eliminates the Curie line. The gas-liquid binodals in presence
of the external magnetic field are depicted with thin lines. The attached
numbers are strengths of the field $h/J_0$. At small $v$ there is a
temperature interval, where the external field suppresses phase separation
--- the top of binodal shifts downward in agreement with the results of
Kawasaki \cite{Kawasaki}.  Nevertheless, at large nonmagnetic attractions
(e.g., $v=4$) the reverse field effect takes place. At moderate values of
the nonmagnetic interaction (for example, $v=2$) the field effect is
nonmonotoneous --- weak fields lower the top of binodal, stronger
fields shift it up.

In Fig.~\ref{mf&tca}b one can see that the TCA, besides quantitative
differences, gives some qualitative corrections to the MFA results.
Within the TCA the Curie line becomes slightly concave,
and the nonmagnetic attraction between particles increases the Curie
temperature.  The latter effect can be justified by the qualitative
arguments. Indeed, the nonmagnetic attraction increases the probability that
a randomly chosen pair of the nearest neighbor sites is occupied.  Since
at this sites particles interact magnetically, the magnetic interaction
becomes more effective, and the Curie temperature increases also. Therefore
the account of density fluctuations in the TCA leads to the dependence of
the Curie temperature on $v$. In the case of the non-compressible fluid
($n=1$) the density fluctuations are absent, and the Curie temperature is
independent of the nonmagnetic attraction.

The TCA predictions concerning the effect of field support the MFA
results.  The variety of field effects may be explained by the existence
of two concurrent tendencies. {\em The first}, the external field aligns
the spins, which leads to the more effective attraction between
particles (let us remind that at $v=0$ particles with parallel spins
($S_iS_j=1$) attract and those with opposite spins ($S_iS_j=-1$) repulse).
This raises the binodal (for example, in simple nonmagnetic fluids the
binodal goes up when the interaction increases). {\em The second} tendency
takes place, if the susceptibility of the rarefied phase is larger than
that of the coexisting dense phase. In this case the magnetization and,
consequently, the effective attraction between particles grow better in
the rarefied phase.  This decreases the energetical gain of the phase
separation.  Therefore the second tendency suppresses the gas-liquid
separation in the fluid and counteracts the first tendency.  The second
tendency is very strong at $h=0$ and $v=0$ in the region of the
tricritical point, where the vapor (paramagnetic) branch of binodal almost
coincides with the Curie line (where the susceptibility tends to
infinity), whereas the branch of the coexistent liquid phase rapidly
deviates from the Curie line. As a result, the external field lowers the
top of the binodal.  The second tendency gets weak and disappears when the
susceptibilities of the coexistent phases levels; they are comparable, for
example, if both liquid and vapor phases are paramagnetic.  The behavior of
$v=4$ binodal (see Fig.~\ref{mf&tca}) demonstrates this feature. A
relation between the susceptibilities results from various factors. For
example, a short-rangeness of the interactions levels the susceptibilities
and weakens the second tendency \cite{PRL}. It can be seen from the
following observation of the field effect at $v=2$:  in
Fig.~\ref{mf&tca}a the top of binodal at $h=0.01$ is higher than that
at $h=0.1$, whereas in Fig.~\ref{mf&tca}b the reverse situation takes
place.  Since the results of the MFA (as well as those of the TCA in the
limit $z\to\infty$) are correct for the long-ranged potentials, whereas
for our case the TCA is much more accurate, the corrections provided by
the TCA have to be attributed to differences between the systems with the
long-range and short-range potentials.

\begin{figure}
\noindent
\epsfxsize=\columnwidth\epsffile{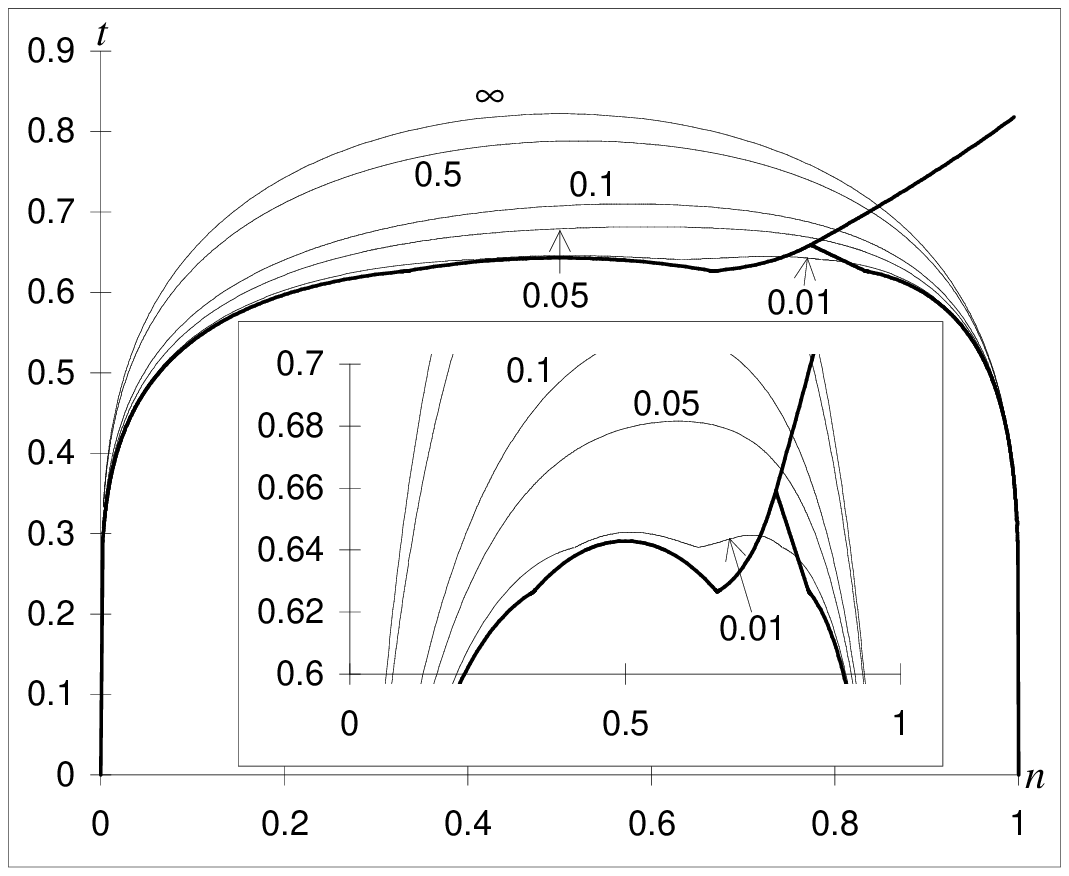}
\caption{The phase diagram of the model at $v=3$ within the TCA. The inset
shows the vicinity of the top of $h=0$ binodal in a more detail.}
\label{triple} \end{figure}

Still more illuminating confirmation of the ``bi-tendency" explanation one
can see in Fig.~\ref{triple}. There is the phase diagram of special
topology, which takes place at intermediate $v$. The model at $h=0$ and
$t=0.63$ undergoes two first-order phase transitions. At this temperature
the fluid can be in three phases: paramagnetic gas (at $n<0.35$),
paramagnetic liquid ($0.65<n<0.69$), ferromagnetic liquid ($n>0.83$). What
we would like to emphasize is that weak external fields (e.g.,
$h/J_0=0.01$) raise the binodal at $n=0.5$ and lower it at $n=0.75$. Such
behavior completely fit into the ``bi-tendency" explanation:  at $n=0.5$
and $h=0$ both phases are paramagnetic, the second tendency is absent,
therefore the external field favors the phase separation; at $n=0.75$ the
second tendency wins at small fields, like in the case $v=2$ (see
Fig.~\ref{mf&tca}).

One can see that the lattice gas approach may be successfully used for
description of complex fluids when continual approaches lead to too
complex calculations or do not give satisfactory results. It is this
situation that takes place when one determines the influence of the
nonmagnetic attraction on the Curie temperature \cite{PRL,GD}. In this
case the account of the short-range correlations within the cluster
approach yields qualitatively new results in comparison with current
continual methods.

\end{multicols}

\end{document}